\documentclass{article}

\usepackage{PRIMEarxiv}

\usepackage[utf8]{inputenc} 
\usepackage[T1]{fontenc}    
\usepackage{hyperref}       
\usepackage{url}            
\usepackage{booktabs}       
\usepackage{amsfonts}       
\usepackage{nicefrac}       
\usepackage{microtype}      
\usepackage{lipsum}
\usepackage{fancyhdr}       
\usepackage{graphicx}       
\usepackage{graphicx}%
\usepackage{multirow}%
\usepackage{amsmath,amssymb,amsfonts}%
\usepackage{amsthm}%
\usepackage{mathrsfs}%
\usepackage[title]{appendix}%
\usepackage{textcomp}%
\usepackage{manyfoot}%
\usepackage{booktabs}%
\usepackage{algorithm}%
\usepackage{algorithmicx}%
\usepackage{algpseudocode}%
\usepackage{listings}%
\usepackage{tabularray}
\usepackage{subcaption}
\usepackage{comment}
\usepackage[table,xcdraw]{xcolor}

\newcommand\MyBox[2]{
	\fbox{\lower0.75cm
		\vbox to 1.7cm{\vfil
			\hbox to 1.7cm{\hfil\parbox{1.4cm}{#1\\#2}\hfil}
			\vfil}%
	}%
}

\title{Machine Learning for Windows Malware Detection and Classification: Methods, Challenges and Ongoing Research
\thanks{This is a preprint of the Chapter to be published in the Book \emph{Malware - Handbook of Prevention and Detection}, Springer.}
}

\author{
  Daniel Gibert\\
  CeADAR, Ireland's Centre for Artificial Intelligence\\
  University College Dublin\\
  Dublin\\
  \texttt{daniel.gibert@ucd.ie} \\
}

\begin{document}
\maketitle

\begin{abstract}
In this chapter, readers will explore how machine learning has been applied to build malware detection systems designed for the Windows operating system. This chapter starts by introducing the main components of a Machine Learning pipeline, highlighting the challenges of collecting and maintaining up-to-date datasets. Following this introduction, various state-of-the-art malware detectors are presented, encompassing both feature-based and deep learning-based detectors. Subsequent sections introduce the primary challenges encountered by machine learning-based malware detectors, including concept drift and adversarial attacks. Lastly, this chapter concludes by providing a brief overview of the ongoing research on adversarial defenses.
\end{abstract}


\section{Building a Machine Learning-based Malware Detector from Scratch: Main Components}
\label{sec:ml_pipeline}
A machine learning (ML) pipeline, independently of the domain application, typically consists of the following key components:

\begin{enumerate}
	\item Data Collection. Data collection is the process of gathering/collecting data for the purposes of training our machine learning-based model. In the context of Windows malware detection, our focus is on collecting binaries including benign and malicious software. This process is crucial as the quality and relevance of the data directly impacts the performance of the model.
	\item Data Preprocessing. Data preprocessing is the process of converting the binaries into a suitable format for analysis. For example, analyzing the file statically or dynamically to extract information from its hexadecimal representation, assembly language source code, file metadata, the sequence of API calls invoked during dynamic analysis, etcetera. This step involves extracting and engineering features, and it is responsible for transforming the raw data into a more suitable format for the machine learning models. 
	\item Model Training and Evaluation. The Model training and evaluation process includes: (1) choosing an appropriate machine learning algorithm such as decision trees, random forests, and neural networks; (2) training the chosen model(s) on the training data; and (3) optimizing the hyperparameters and validating the model's performance on the validation set to ensure the model generalizes well to unseen data.
	\item Model Deployment, Monitoring and Maintenance. Model deployment is the process of deploying the model for its intended use. This typically involves the integration of the trained model into a system for real-time malware detection. Model monitoring and maintenance involves monitoring the model's performance in the deployed environment to detect any degradation or drift in performance and to regularly update the model with new benign and malicious examples to adapt to new patterns and maintain its effectiveness~\cite{10102612}. The readers are referred to Section~\ref{sec:concept_drift} for a more detailed description of the methods developed to address concept drift.
\end{enumerate}
Now, we will take a look at each of these components in more detail.
\subsection{Data Collection}
Data collection is the process of gathering/collecting data relevant to the Machine Learning project's goals and objectives. This process is the first and most fundamental step in a machine learning pipeline as it directly impacts the performance of the ML-based model. In the machine learning field, there is a good old principle that says: "\emph{garbage in, garbage out}", which means that a machine learning model will learn exactly what it's taught. This principle highlights the critical importance of the quality of the input data in determining the quality of the output generated by the model. 

If the data used for training a machine learning model is of poor quality, inaccurate, incomplete, or biased, the resulting model's performance will likely reflect these shortcomings, regardless of how complex and sophisticated the model is, how skilled the data scientists are, or how much money have been spent on the project. The model can only learn patterns based on the information it is given during training. Therefore, it is of critical importance to rigorously collect and verify the data before feeding it into the machine learning models.

In the context of malware detection, data collection involves gathering and labeling both benign and malicious executables from several sources. Windows software, especially commercial software, is protected by copyright laws. As a result, distributing software openly without proper authorization from copyright holders can lead to legal issues.
To circumvent this limitation, researchers typically collect the benign data from clean installations of various versions of the Windows OS, i.e. Windows XP, Windows 7, Windows 10, etcetera, and from software package managers such as Chocolatey~\footnote{\url{https://chocolatey.org/}} or Windget~\footnote{\url{https://github.com/microsoft/winget-cli}}, among others. In addition, researchers usually provide the hash of the executables used in their experiments to facilitate their reproduction\cite{2018arXiv180404637A,harang2020sorel20m,bodmas}. Additionally, malware can be collected from multiple sources as follows:
\begin{itemize}
	\item Malware repositories such as VirusTotal~\footnote{\url{https://virustotal.com/}}, MalShare~\footnote{\url{https://malshare.com/}}, VirusShare~\footnote{\url{https://virusshare.com/}}, among others. These repositories provide access to a vast collection of malware samples which are updated daily. 
	\item Threat Intelligence feeds. Cybersecurity companies usually collect and analyze malware for research and for business purposes. Some of them share threat intelligence feeds containing information about known malware samples or recently analyzed malware.
	\item Honeypots. Another option to capture fresh malware samples is to set up a honeypot.
	\item Dark web. One can also monitor the dark web to look for new or customized malware. 
\end{itemize}

\subsubsection{Datasets} 
The availability of public datasets has played a significant role in advancing scientific knowledge. Public datasets are crucial for the following reasons: (1) they allow researchers to reproduce and validate the results of previous studies; (2) they facilitate benchmarking the performance of algorithms, models and methods, fostering competition and promoting the development of better solutions; and (3) they save costs and time. In addition, in the context of Windows malware detection, due to the dynamic evolving nature of benign and malicious software, there is a constant need for up-to-date datasets, including new malware families and variants that appear over time. Table~\ref{tab:windows_datasets} presents an overview of the available datasets most used in the literature. For a more detailed description of the datasets the readers are referred to the original papers~\cite{DBLP:journals/corr/abs-1802-10135,2018arXiv180404637A,harang2020sorel20m,bodmas}.

\begin{table}
	\centering
	\begin{talltblr}[
		caption={Summary of the benchmarks for Windows malware research.},
		label={tab:windows_datasets}
		]{
			width = \linewidth,
			colspec = {Q[140]Q[140]Q[100]Q[90]Q[96]Q[90]Q[163]Q[131]},
			hline{1-2,6} = {-}{},
		}
		& Timestamp         & \#Samples  & \#Benign  & \#Malware & \#Families & Malware Binaries   & Feature Vectors \\
		Microsoft\cite{DBLP:journals/corr/abs-1802-10135} & Before 2015       & 10,868     & 0         & 10,868      & 9          & Without PE headers & No              \\
		EMBER\cite{2018arXiv180404637A}     & 2017-2018 & 2,050,000  & 750,000   & 800,000     & N/A        & No                 & Yes             \\
		SOREL-20M\cite{harang2020sorel20m} & 2017-2019 & 19,724,997 & 9,762,177 & 9,962,820   & N/A        & Yes                & Yes             \\
		BODMAS\cite{bodmas}   & 2019-2020 & 134,435    & 77,142    & 57,293      & 581        & Yes                & Yes             
	\end{talltblr}
\end{table}

\subsection{Data Preprocessing}
Data preprocessing is a crucial step necessary to prepare the data for the machine learning-based malware detectors. It involves cleaning and transforming the data to make it suitable for analysis. Approaches can be grouped into (1) static analysis and (2) dynamic analysis approaches. 

Static analysis refers to the analysis of the computer programs without running them. This type of analysis involves examining the PE structure, including its headers, the dynamic library references for linking, the sections, etcetera. Furthermore, programs can be disassembled to obtain their corresponding assembly language source code, which provides insights about low-level instructions that the program would execute and the interactions of the program with memory. In contrast, dynamic analysis involves executing the program in a safe environment and monitoring its execution. Dynamic analysis is typically employed when static analysis has reached an impasse, be it due to obfuscation, packing, or when the analysts have depleted the available static analysis techniques. Unlike static analysis, dynamic analysis lets you observe the malware’s true functionality. Depending on the type of analysis, machine learning-based models are classified into static and dynamic detectors.

\subsection{Model Training and Model Evaluation}
The Model training and evaluation process includes (1) selecting the most appropriate machine learning algorithm, (2) training, and (3) evaluating the model:
\begin{itemize}
	\item Model selection. This component of the pipeline involves choosing the most appropriate machine learning algorithm for the task at hand. The characteristics of the input data will determine the type of algorithm that will be most suitable. For instance, if the input data is a sequence of bytes, convolutional neural networks will be more suitable than a Support Vector Machine~\cite{DBLP:conf/aaai/RaffBSBCN18} as convolutional neural networks are designed to capture spatial relationships within the input data. On the other hand, if the input data is a set of features extracted from PE files, then a decision tree might be more suitable than a neural network~\cite{2018arXiv180404637A}, as they are inherently more interpretable.
	\item Training the model. This component involves training the chosen model(s) on the training dataset, and optimizing its hyperparameters.
	\item Model evaluation. This component involves validating the model's performance using the validation set to fine-tune its parameters and ensure it generalizes well to unseen data. In the context of malware detection, the system predicts whether a given input file is benign or malicious, and thus, the four following outcomes can occur, as illustrated in Table~\ref{tab:confusion_matrix}:
	\begin{itemize}
		\item True positive. The file is malware and the model says it is malware.
		\item False negative. The file is malware and the model says it is benign.
		\item False positive. The file is benign and the model says it is malware.
		\item True negative. The file is benign and the model says it is benign.
	\end{itemize}
\end{itemize}

\begin{table}[ht]
	\centering
	\caption{Confusion matrix illustrating the four possible outcomes of a malware detection system.}
	\label{tab:confusion_matrix}
	\noindent
	\renewcommand\arraystretch{1.5}
	\setlength\tabcolsep{0pt}
	\begin{tabular}{c >{\bfseries}r @{\hspace{0.7em}}c @{\hspace{0.4em}}c @{\hspace{0.7em}}l}
		\multirow{10}{*}{\rotatebox{90}{\parbox{1.1cm}{\bfseries\centering actual\\ value}}} & 
		& \multicolumn{2}{c}{\bfseries Prediction outcome} & \\
		& & \bfseries p & \bfseries n & \bfseries total \\
		& p$'$ & \MyBox{True}{Positive (TP)} & \MyBox{False}{Negative (FN)} & P$'$ \\[2.4em]
		& n$'$ & \MyBox{False}{Positive (FP)} & \MyBox{True}{Negative (TN)} & N$'$ \\
		& total & P & N &
	\end{tabular}
\end{table}

While achieving high accuracy is generally desirable, it may not be sufficient on its own to evaluate the system's performance, especially in imbalanced datasets where one class (e.g., benign software) significantly dominates the other (e.g., malware). In such scenarios, relying solely on accuracy can be misleading, as the classifier may achieve high accuracy by simply predicting the majority class most of the time. In these situations metrics like the $F1$ score become crucial. The $F1$ score takes into account both precision and recall, providing a more balanced assessment of a model's performance. Precision
measures the proportion of positive identifications that are actually correct, i.e. $\text{Precision} = \frac{TP}{TP+FP}$. Recall measures the proportion of actual positive instances that were correctly identified by the model, i.e. $\text{Recall} = \frac{TP}{TP+FN}$. The $F1$ score is calculated using the following Equation:
\begin{equation}
	F1 = 2*\frac{\text{Precision}\, \times  \, \text{Recall}}{\text{Precision}\, +\, \text{Recall}}
\end{equation}

\section{Static Machine Learning-based Malware Detectors}
\label{sec:static_detectors}
Static machine learning-based malware detectors are a type of malware detectors designed to determine if a given piece of software is benign or malicious from information extracted using static analysis techniques. Unlike dynamic analysis, which involves running the code in a controlled environment to observe its behavior, static analysis techniques examine the code itself, its structure, patterns and characteristics to determine if the program is potentially malicious. 

\begin{figure}[ht]
	\includegraphics[width=0.3\columnwidth]{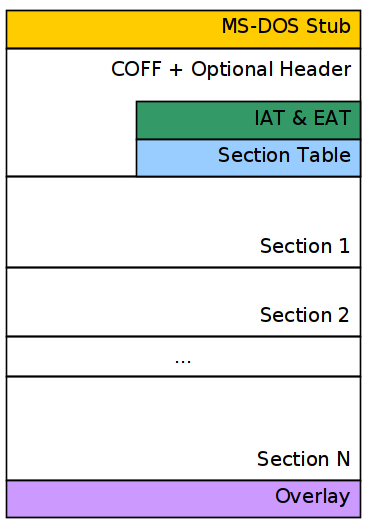}
	\centering
	\caption{A graphical depiction of the PE file format.}
	\label{fig:pe_file_structure}
\end{figure}

The Windows operating system uses the Portable Executable (PE) file format to represent programs, executables, and DLLs (Dynamic Link Libraries). The PE file comprises multiple headers and sections, as illustrated in Figure~\ref{fig:pe_file_structure}, that provide directions to the dynamic linker to map the file into memory. A PE file consists of:
\begin{itemize}
	\item MS-DOS Stub. The MS-DOS stub is a small executable program that is included in the headers (first 64 bytes) that typically displays a message when the PE file is run on MS-DOS or in a DOS-like environment. In modern Windows systems, the DOS stub is ignored, but it remains as part of the PE file structure for compatibility reasons.
	\item COFF header. The COFF header contains information about the type of target machine, the number of sections, a timestamp indicating where the file has been created, the file offset of the COFF symbol table, the number of entries in the symbol table, the size of the Optional header, and attributes of the file.
	\item Optional header. The Optional header contains additional information that is required by the linker and loader in Windows, including the Import Address Table (IAT) and the Export Address Table (EAT) addresses and sizes, and the Sections Table.
	\begin{itemize}
		\item Import Address Table. The Import Address Table is a data structure that contains the addresses of functions imported by the program at runtime.
		\item Export Address Table. The Export Address Table is a data structure that contains the addresses of functions that the program exports for use by other executables.
		\item Section Table. The Sections Table contains information about the various sections present in the executable file, such as code sections, data sections, resource sections, and others. Each entry in the Section Table describes a specific section's characteristics, including its name, size, memory location, and characteristics (such as executable code or read-only data).
	\end{itemize}
	\item Sections. The PE headers are followed by the actual data and content of the sections in the PE file. The sections may contain executable code, data, resources, and other information that makes up the program. The layout of sections and their characteristics is specified in the Sections Table. 
\end{itemize}

To access the information within a PE file, researchers can use the Pefile~\footnote{\url{https://pypi.org/project/pefile/}} and LIEF~\footnote{\url{https://pypi.org/project/lief/}} libraries. Both libraries provide a straightforward way to parse the headers, retrieve attributes, read the import address table (IAT) and export address table (EAT), and the sections table. In addition, LIEF allows users to modify the PE files by injecting new sections, importing new libraries and functions, etcetera.

Furthermore, researchers can obtain the assembly language source code from a PE file through a process called disassembly. Disassembly involves converting the machine code into human-readable assembly language instructions. The assembly language source code provides insights about the low-level instructions that the program would execute, and how the program interacts with memory and the Windows API library. There are various free and premium disassemblers that researchers can use to disassemble a PE file. The most well-known are IDA Pro~\footnote{\url{https://hex-rays.com/ida-pro/}}, Ghidra~\footnote{\url{https://ghidra-sre.org/}} and Radare2~\footnote{\url{https://rada.re/n/}}.

To sum up, researchers can extract a wide range of information from both the PE file and its assembly language code representation to build their static ML-based malware detectors. Depending on the input, static ML-based malware detectors are grouped into (1) feature-based detectors, (2) deep learning-based detectors, and (3) hybrid detectors.

\subsection{Feature-based Detectors}
Feature-based detectors~\cite{10.1145/2857705.2857713,2018arXiv180404637A} rely on extracting specific features or characteristics from PE files or their corresponding assembly language source code that are believed to be indicative of malware. Typically, feature-based detectors extract one or more types of features and then concatenate the extracted feature vectors into a single feature vector that is used to represent the computer program. Features commonly extracted are byte and opcode N-grams, file metadata, API libraries and functions usage, structural features, among others. Next, I briefly describe two state-of-the-art approaches for malware detection and classification that extract various types of features from the PE files and their corresponding assembly language source code.

\subsubsection{EMBER LightGBM Model}
Anderson et al.~\cite{2018arXiv180404637A} introduced the EMBER dataset, a labeled benchmark dataset for training static machine learning-based malware detectors. The EMBER dataset includes eight groups of features extracted from 1.1M binary files. In addition, they provided a baseline gradient boosted decision tree model trained using LightGBM. Following you will find a brief description of each group of features:
\begin{itemize}
	\item General file information. This set of features includes the file size and basic information obtained from the PE header such as the virtual size of the file, the number of imported and exported functions, among others.
	\item Header information. This group of features includes features extracted from the COFF header and the Optional header, including the timestamp, the target machine, a list of image characteristics, the target subsystem, DLL characteristics, the file magic as a string, major and minor image versions, linked version, and system and subsystem versions.
	\item Imported functions. This feature set refers to the API libraries and functions imported by the PE file vectorized using the hashing trick. This information can be extracted from the Import Address Table. 
	\item Exported functions. This set of features corresponds to the list of exported functions, extracted from the Export Address Table, summarized using the hashing trick.
	\item Section information. This group of features summarizes the properties of each section, extracted from the Sections Table, including the name, size, entropy, virtual size, and a list of strings representing the Section characteristics.
	\item Byte histogram. The byte histogram refers to the frequency of each byte within the executable.
	\item Byte-entropy histogram~\cite{7413680}. This feature set is calculated by sliding a 2048 byte window over the file with a step size of 1024 bytes, computing the entropy of the bytes within the window and pairing it with each byte occurrence within the window.
	\item String information. This set of features consists of simple statistics about printable strings that are at least five characters long including the number of strings, their average length, a histogram of the printable characters within those strings, and the entropy of characters across all printable strings.
\end{itemize}

\subsubsection{Novel Feature Extraction, Selection and Fusion for Effective Malware Family Classification}
Ahmadi et al.~\cite{DBLP:journals/corr/abs-1802-10135} presented a malware classification framework that extracts multiple types of features from both the hexadecimal representation and the assembly language source code of computer programs, as summarized in Table~\ref{tab:feature_fusion}. Then, the most relevant types of features are selected and stacked into a single vector using a modified version of the forward stepwise selection technique. This feature selection technique can be summarized as follows: starting with a feature vector with no features, it gradually augments the feature vector by adding a single type of features. At each step, the feature vector that produces the minimum value of the multiclass logarithmic loss will be added to the final feature vector. The process stops when adding more features does not decrease the value of loss. Afterwards, the resulting feature vector consisting of the most relevant types of features is used to train a malware classifier using XGBoost~\cite{10.1145/2939672.2939785}.

\begin{table}[ht]
	\centering
	\begin{talltblr}[
		caption={Summary of the features extracted in Ahmadi et al.~\cite{DBLP:journals/corr/abs-1802-10135}.},
		label={tab:feature_fusion}
		]{
			width = \linewidth,
			colspec = {Q[167]Q[123]Q[650]},
			cell{2}{1} = {r=5}{},
			cell{7}{1} = {r=8}{},
			hline{1-2,7,15} = {-}{},
		}
		File Type                     & Feature Type           & Description                                                                                                                                                 \\
		PE file                       & Byte unigrams            & Frequency of each byte.                                                                                                                                      \\
		& Metadata               & Size of the file, and address of the first bytes sequence.                                                                                                   \\
		& Entropy statistics     & Quantiles, percentiles, mean and variance of the structural entropy representation.                                                                          \\
		& Image-based features   & Haralick and Local Binary Pattern features extracted from the executables' grayscale image representation.                                                   \\
		& String-based histogram & Histograms related to the distribution of length of ASCII strings found in the PE file.                                                                      \\
		Assembly Language Source Code & Metadata               & Size of the file, and the number of lines in the file.                                                                                                       \\
		& Symbols frequency      & Frequencies of the following set of symbols: \{-, +, *, [, ], ?, @\}.                                                                                        \\
		& Opcode unigrams         & Frequency of a subset of operation codes.                                                                                                                            \\
		& Register frequency     & Frequency of use of the registers.                                                                                                                           \\
		& API calls              & Frequency of use of Windows Application Programming Interface (API) libraries and functions.                                                                                         \\
		& Section-based features & Features extracted from the sections such as the total number of .text, .data, .idata sections, the proportion of known sections to all sections, etcerera. \\
		& Data define features   & Features related to the data define db, dw, and dd instructions.                                                                                            \\
		& Miscellaneous          & Frequency of 95 manually chosen keywords.                                                                                                                    
	\end{talltblr}
\end{table}

\subsection{Deep Learning-based Detectors}
Deep learning-based detectors use deep neural networks to automatically learn intricate patterns and representations directly from the raw data without explicit feature engineering. Deep learning-based detectors replace the feature engineering step with a convolutional neural network with one or more convolutional layers that learn patterns from benign and malicious executables represented as a sequence of bytes (See Section~\ref{sec:byte_based_detectors}) or a sequence of assembly language instructions (See Section~\ref{sec:assembly_based_detectors}). Moreover, deep neural networks have also been used to classify malware into families based on some sort of visual representation (Section~\ref{sec:visualization_techniques})

\subsubsection{Byte-based Detectors}
\label{sec:byte_based_detectors}
Byte-based detectors, commonly referred to as end-to-end malware detectors~\cite{DBLP:conf/aaai/RaffBSBCN18,DBLP:conf/iclr/KrcalSBJ18} directly process the sequence of bytes present in a file instead of relying on manual feature engineering. As a result, end-to-end detectors eliminate the need for expert-crafted features, automating the feature extraction and detection process. However, this implies treating each byte as a unit in a sequence, and thus, the problem of malware detection turns out to be a sequence classification problem on the order of millions of time steps. As a result, the deep learning architectures that have been proposed so far have been limited in terms of complexity and depth, and typically consist of shallow architectures with convolutional layers with big filter sizes and strides followed by a global max pooling or global average pooling layer to reduce the computational requirements.

\begin{figure}[ht]
	\includegraphics[width=0.7\columnwidth]{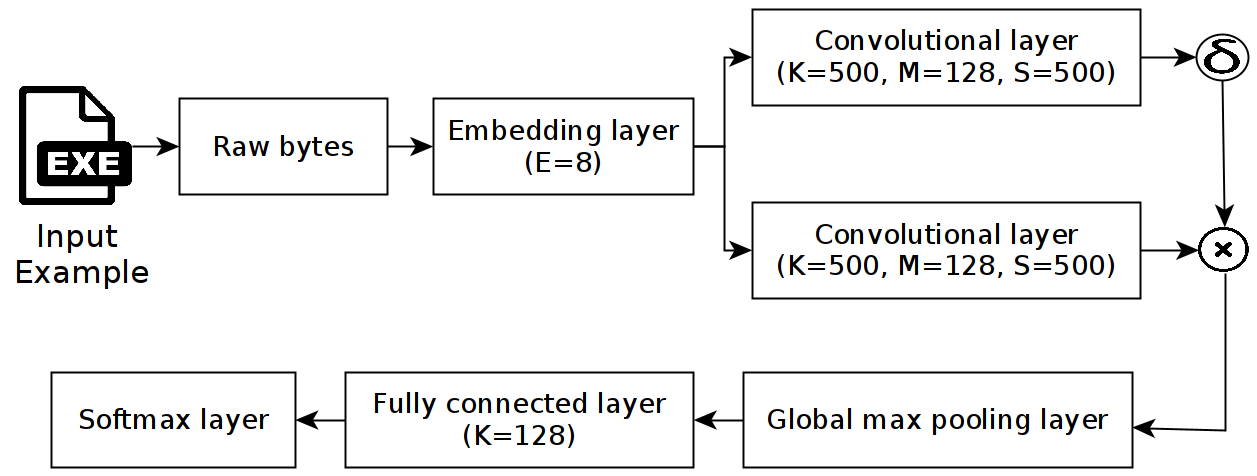}
	\centering
	\caption{MalConv architecture~\cite{DBLP:conf/aaai/RaffBSBCN18} .}
	\label{fig:malconv_architecture}
\end{figure}

Raff et al.~\cite{DBLP:conf/aaai/RaffBSBCN18} proposed MalConv, a shallow convolutional neural network to detect malware based on raw byte sequences. The MalConv architecture consists of an embedding layer, a gated convolutional layer, a global-max pooling layer and a fully-connected layer. Cf. Figure~\ref{fig:malconv_architecture}.

\begin{figure}[ht]
	\includegraphics[width=0.7\columnwidth]{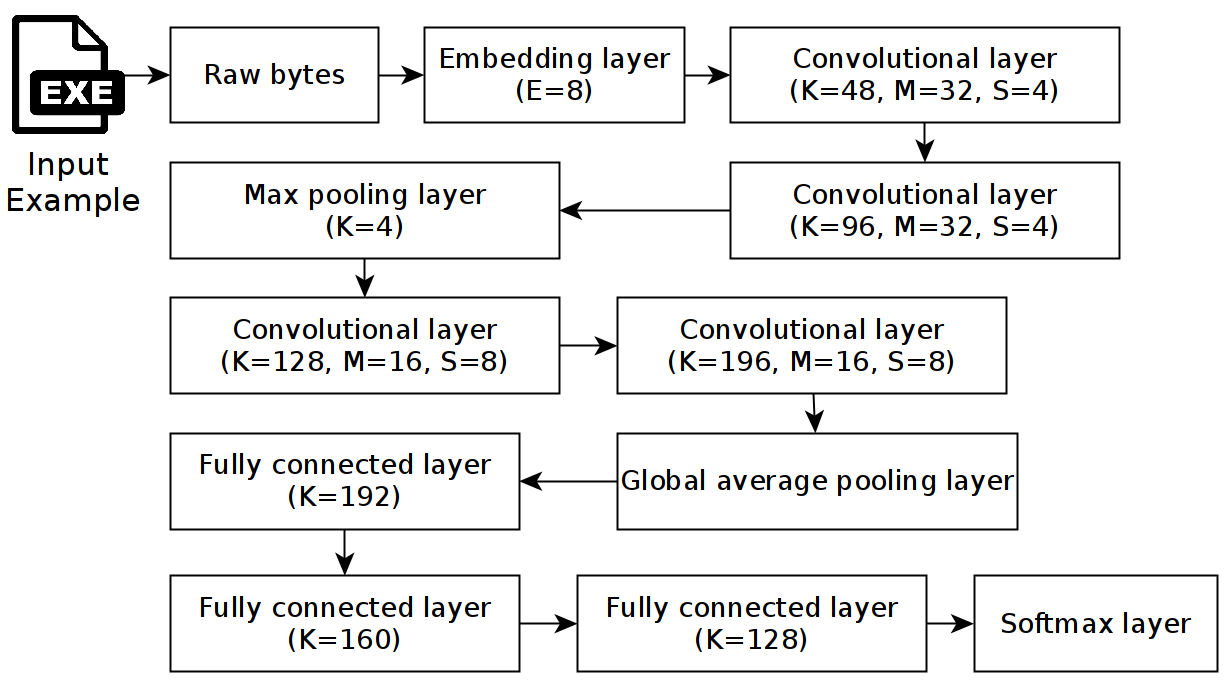}
	\centering
	\caption{AvastConv architecture.}
	\label{fig:avastconv_architecture}
\end{figure}

Krcal et al.~\cite{DBLP:conf/iclr/KrcalSBJ18} proposed AvastConv, a deep learning architecture that comprises four convolutional layers with kernel sizes of 32, 32, 16, and 16, respectively. The 2nd and 3rd convolutional layers are separated by a max pooling layer. Furthermore, a global average pooling layer is located after the 4th convolutional layer. Subsequently, the architecture includes three fully connected layers with sizes of 192, 192, and 160, culminating in a softmax layer. Cf.Figure~\ref{fig:avastconv_architecture}.

\subsubsection{Assembly Language Instructions-based Detectors}
\label{sec:assembly_based_detectors}
\begin{figure}[ht]
	\includegraphics[width=0.9\columnwidth]{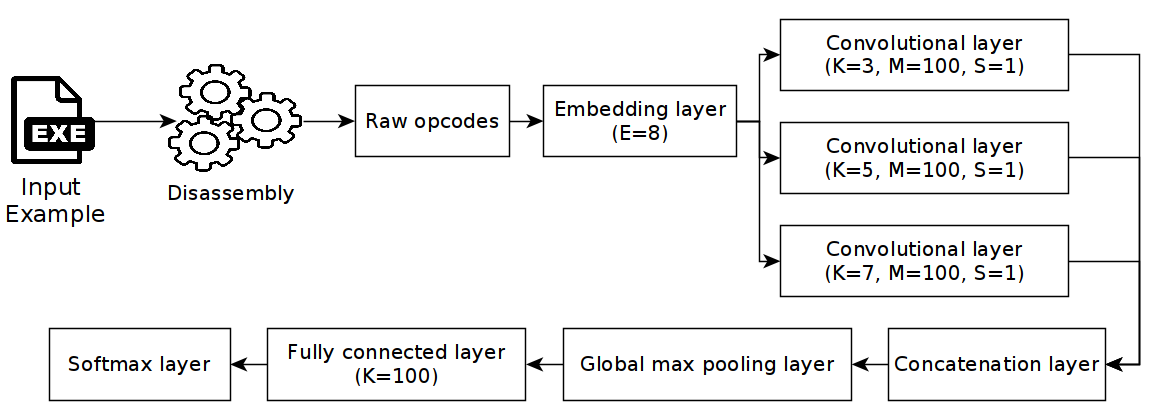}
	\centering
	\caption{ShallowConv architecture~\cite{DBLP:conf/ccia/GibertBMPSV17,GIBERT2021102159}.}
	\label{fig:shallowconv_architecture}
\end{figure}

Similarly, Gibert el al.~\cite{DBLP:conf/ccia/GibertBMPSV17,GIBERT2021102159} presented a shallow convolutional neural network to learn N-gram like features from the operation codes, also known as opcodes, found in the assembly language source code of computer programs. Cf. Figure~\ref{fig:shallowconv_architecture}. An N-gram is a contiguous sequence of N items from a given sequence of text. This architecture takes as input the opcodes of the assembly language instructions extracted from the assembly language source code of the computer programs. An opcode is the portion of the machine language instruction that specifies the operation to be performed, i.e. \emph{add}, \emph{sub}, \emph{jmp}, etcetera. Afterwards, every opcode is represented as a low-dimensional vector of real values or word embedding. The rationale behind using word embeddings is to better capture the meaning of the opcodes. After the embedding layer, the architecture has a convolutional layer, a global max-pooling layer, and a fully-connected layer. The convolutional layer applies filters of sizes 3, 5, 7 to the embedded opcodes, allowing the network to extract higher-level features by convolving across the embedded input. These filters can detect specific patterns or N-gram-like features within the sequence of opcodes.
Following the convolutional layer, the global max-pooling layer captures the most salient features from each feature map generated by the convolution. Afterwards, the fully connected layer takes the output from the pooling layer, processes it, and produces the final output of the network, which is converted to probability scores across multiple classes, using a softmax function.

\subsection{Visualization Techniques}
\label{sec:visualization_techniques}
In an attempt to aid security analysts in understanding and analyzing malicious software, various visualization techniques have been proposed to visualize and classify malware into families. The rationale behind using visualization techniques is that, malware binaries belonging to the same family tend to have similar structure and layout while being distinct to malware binaries belonging to a different family. In addition, the visualization techniques do not require to disassemble nor to execute the code of the binaries, making them a good option to complement static detectors.

\subsubsection{Grayscale Image Representation}
\label{sec:grayscale_img_representation}
\begin{figure}[ht]
	\centering
	\includegraphics[width=0.8\columnwidth]{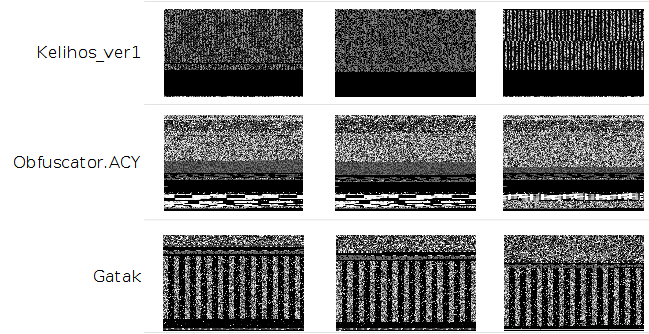}
	\caption{Grayscale image representation of malware binaries belonging to the Kelihos\_ver1, Obfuscator.ACY and Gatak families, respectively~\cite{DBLP:journals/virology/GibertMPV19}.}
	\label{fig:grayscale_img_representation_of_mlw_binaries}
\end{figure}

L. Nataraj et al.~\cite{DBLP:conf/vizsec/NatarajKJM11} was the first work to propose to visualize malware binaries as grayscale images, based on the observation that malware binaries belonging to the same family have a similar layout and texture. Furthermore, as malware authors typically change small parts of the original code to produce new variants, the grayscale images can capture these small changes while retaining the global structure. Cf. Figure~\ref{fig:grayscale_img_representation_of_mlw_binaries}. 

\begin{figure}[ht]
	\centering
	\includegraphics[width=0.7\columnwidth]{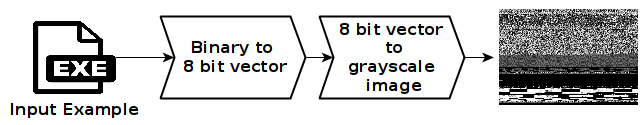}
	\caption{Visualizing malware as a grayscale image~\cite{DBLP:conf/vizsec/NatarajKJM11}.}
	\label{fig:grayscale_visualization_process}
\end{figure}

The process of converting a malware binary to a grayscale image, illustrated in Figure~\ref{fig:grayscale_visualization_process}, is as follows: (1) a malware binary is read as a vector of 8 bit unsigned integers; (2) the resulting vector is reorganized into a 2D array which is then visualized as a grayscale image in the range [0,255] (0: black, 255: white).

 Motivated by this visual similarity, Nataraj et al.~\cite{DBLP:conf/vizsec/NatarajKJM11} proposed to extract GIST features~\cite{gist_features_paper} based on the grayscale images to characterize and classify malware into families using the K-nearest neighborhood algorithm. Due to the simplicity of the grayscale visualization, several studies have expanded on Nataraj et al's work by incorporating various image-based features such as Haralick features~\cite{10.1145/2857705.2857713}, Local Binary Pattern features~\cite{10.1145/2857705.2857713}, PCA features~\cite{7856826}, among others. Furthermore, researchers have explored alternative supervised learning algorithms to replace the K-Nearest Neighborhood algorithm such as feed-forward neural networks~\cite{7856826}, decision trees~\cite{10.1145/2857705.2857713}, support vector machines~\cite{7856826}, and more. 

More recently, reflecting recent trends in the Computer Vision domain, Gibert et al.~\cite{DBLP:journals/virology/GibertMPV19} proposed replacing the traditional feature engineering and classification process with an end-to-end approach. In their work, they advocated the use of convolutional neural networks (CNNs) for feature extraction and classification, as CNNs are particularly well-suited to learning hierarchical representations directly from raw pixel values, eliminating the need for manual feature extraction. Building upon their work, subsequent studies have further advanced the detection capabilities of grayscale image-based malware classifiers by employing more sophisticated neural network architectures such as ResNet~\cite{8260773,DBLP:journals/virology/KhanZK19}, EfficientNet~\cite{CHAGANTI2022103306}, GoogleNet~\cite{DBLP:journals/virology/KhanZK19}, and more recently, Vision Transformers~\cite{10.1145/3607720.3607781}.

\subsubsection{Structural Entropy Representation}
\label{sec:structural_entropy_representation}
\begin{figure}[ht]
	\centering
	\includegraphics[width=0.6\columnwidth]{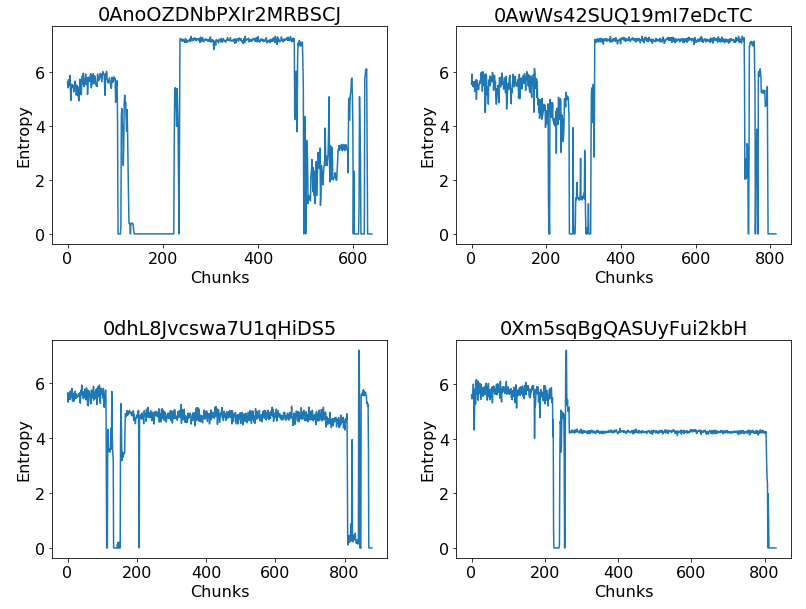}
	\caption{Structural entropy representation of malware binaries~\cite{DBLP:conf/aaai/GibertMPV18}.}
	\label{fig:structural_entropy_mlw_binaries}
\end{figure}

An alternative way to visualize binaries to estimate their similarity is using their structural entropy~\cite{DBLP:conf/aaai/GibertMPV18}. Entropy has long been used to detect the presence of encrypted and compressed segments of code as they tend to have higher entropy compared to native code~\cite{DBLP:journals/ieeesp/LydaH07}. The use of simple entropy statistics, however, is not enough to detect sophisticated malware as malware authors usually try to conceal the encrypted and compressed code in order to bypass entropy filters. On the other hand, by visualizing malware using their structural entropy, analysts can easily detect what sections of the code have been encrypted or compressed. The structural entropy of a binary file is calculated as follows: (1) the binary file is split into non-overlapping chunks of bytes of fixed sizes; (2) for each chunk of bytes the entropy is calculated. This will result in a time series $m=\{m_{0}, m_{1}, m_{n}\}$, where $m_i$ represents the entropy of the $i$-th chunk of bytes in a given binary file. Similarly to the grayscale representation, and as shown in Figure~\ref{fig:structural_entropy_mlw_binaries}, malware binaries belonging to the same family have similar layout. Motivated by the visual similarity of the structural entropy of malware binaries belonging to the same family, Gibert et al.~\cite{DBLP:conf/aaai/GibertMPV18} proposed an end-to-end approach to classify malware into families using convolutional neural networks, showing comparable detection accuracy to K-Nearest Neighbor using the Dynamic Time Warping metric instead of the Euclidean distance.

\subsection{Hybrid Detectors}
Hybrid detectors, in the context of malware detection, refer to those detectors that integrate both traditional feature engineering techniques and deep learning. On the one hand, feature engineering involves manually crafting features that capture the distinct characteristics of malware. On the other hand, deep learning approaches are capable of automatically learning complex patterns from raw data, eliminating the need for extensive manual feature engineering. By using both feature engineering and deep learning, hybrid detectors aim to harness the strengths of both types of approaches to increase their detection capabilities. Following are described two hybrid approaches~\cite{GIBERT2020101873,GIBERT2022117957} that achieve SOTA accuracy on the Microsoft Malware Classification Challenge dataset~\cite{DBLP:journals/corr/abs-1802-10135}.

\begin{figure}[ht]
	\centering
	\includegraphics[width=0.8\columnwidth]{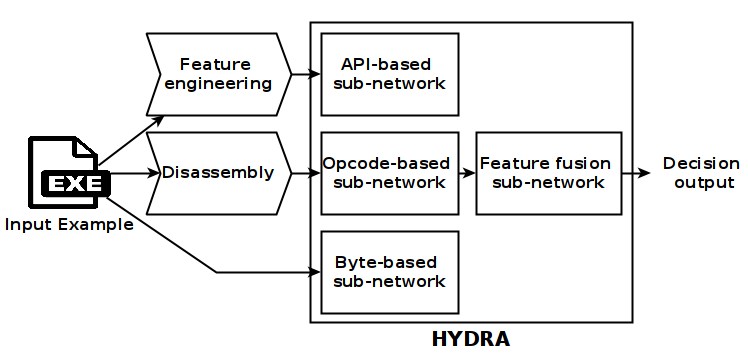}
	\caption{HYDRA architecture~\cite{GIBERT2020101873}.}
	\label{fig:hydra_architecture}
\end{figure}

HYDRA~\cite{GIBERT2020101873}, illustrated in Figure~\ref{fig:hydra_architecture}, is a hybrid malware detector that consists of four sub-networks or components:
\begin{itemize}
	\item A feed-forward neural network that processes the API-based features extracted from malware binaries.
	\item A shallow convolutional network similar to Gibert et al's ShallowConv~\cite{DBLP:conf/ccia/GibertBMPSV17} that extracts opcode N-gram like features from malware's assembly language source code.
	\item A deep convolutional neural network based on AvastConv~\cite{DBLP:conf/iclr/KrcalSBJ18} that extracts byte-based features from binaries represented as a sequence of bytes.
	\item A feed-forward neural network that processes, and combines the features learned by each sub-network, to produce the final decision output.
\end{itemize}
By integrating both traditional features such as API-based features with deep learning, HYDRA achieved comparable detection accuracy to SOTA detectors~\cite{DBLP:journals/corr/abs-1802-10135} by only using as input three modalities of information, i.e. API-based features, opcodes and bytes sequences.

\begin{figure}[ht]
	\centering
	\includegraphics[width=1.0\columnwidth]{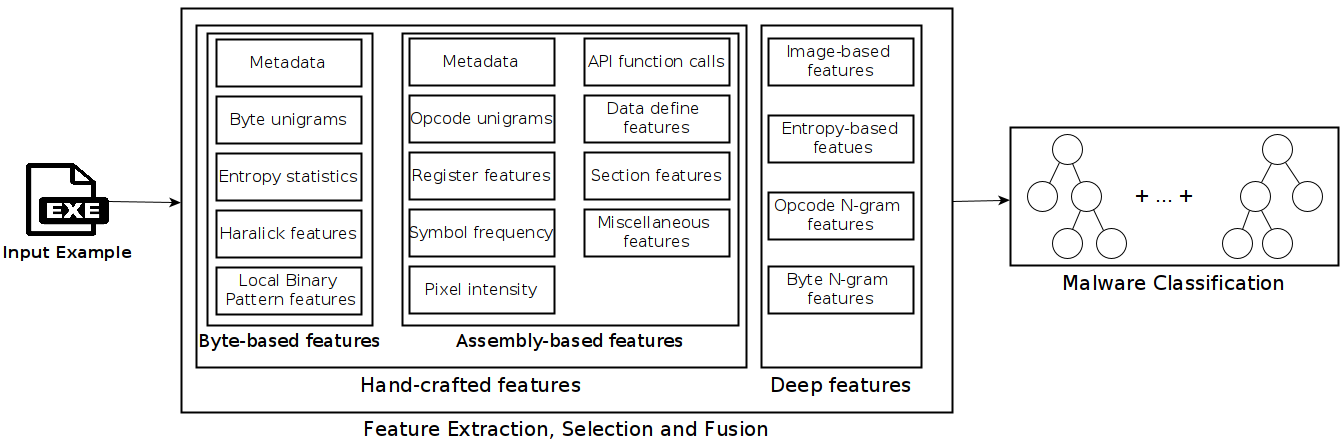}
	\caption{Hybrid malware classification system overview~\cite{GIBERT2022117957}.}
	\label{fig:hybrid_malware_classification_early_fusion}
\end{figure}

Similarly, Gibert et al.~\cite{GIBERT2022117957} proposed a multimodal malware classification system, shown in Figure~\ref{fig:hybrid_malware_classification_early_fusion}, that uses hand-crafted features and deep features, i.e. features extracted by deep learning algorithms. This system extracts well-known hand-crafted features~\cite{DBLP:journals/corr/abs-1802-10135} and it combines these features with what are known as deep learning features, i.e. features that are automatically extracted from raw data using deep learning techniques, typically through the layers of a deep neural network. In their work, Gibert et al.~\cite{DBLP:conf/ccia/GibertBMPSV17,GIBERT2021102159} extracted byte and opcode N-gram like features as well as deep features from malware represented as grayscale images~\cite{DBLP:conf/vizsec/NatarajKJM11,DBLP:journals/virology/GibertMPV19} and its structural entropy representation~\cite{DBLP:conf/aaai/GibertMPV18}. Afterwards, all the extracted features are combined into a single feature vector that is used to characterize and classify malware using gradient boosting trees. 

\section{Open Challenges}
While machine learning has shown promising results in improving malware detection capabilities, several challenges persist~\cite{10102612}, being concept drift and the vulnerabilities of machine learning models to adversarial attacks, the two most urgent challenges. Addressing these challenges is essential to improving the reliability and robustness of ML-based malware detection systems deployed in production.


\subsection{Concept Drift}
\label{sec:concept_drift}
Concept drift refers to the problem of the changing underlying relationships in the data. In supervised learning, a machine learning model approximates a mapping function ($f$) given input data ($X$) to predict a target variable ($y$), $y = f(X)$.
In many problems, this mapping is assumed to be static, meaning that the relationships between input and output do not change over time, and thus, the mapping learned from historical data will be valid in the future on new data. However, this is not true for the problem of malware detection and classification.

Software programs, including malware, naturally evolve over time. Software programs are constantly updated to meet new business and user requirements by adding and removing features, to address bugs and issues identified in the software, to optimize the code, etcetera. These changes are expected to be introduced relatively infrequently, but over time, the similarity between previous and future versions of the software are expected to degrade slowly (gradual degradation), with few exceptions, as to when the code base undergoes significant refactoring (sudden degradation). To make things worse, malware is constantly pushed to evolve in order to evade detection and operate. As a result, malware authors are well-motivated to actively try to circumvent detection by employing a wide range of obfuscation strategies~\cite{5975134}. As a result, the prediction quality of the machine learning model is expected to slowly degrade over time~\cite{10190715,DBLP:conf/uss/PendleburyPJKC19}, as new variants and new malware families appear~\cite{DBLP:conf/ccs/ChowKLCAP23}.

To address concept drift, recent studies have proposed two directions: (1)  proactively detecting and rejecting drifting samples~\cite{DBLP:conf/sp/BarberoPPC22} and (2) to proactively detect aging models~\cite{DBLP:conf/uss/JordaneySDWPNC17} to retrain them. The first approach involves identifying instances that differ from the training data distribution and are likely to be misclassified, and reject them. The second approach involves detecting aging models in their early stages, before the model's performance starts to gradually decline over time. By identifying aging models in a timely manner, researchers can retrain the model with the most recent data~\cite{10190715}, ensuring that the model remains effective.

\subsection{Adversarial Attacks}
\label{sec:adversarial_attacks}

\begin{table}[ht]
	\centering
	\begin{talltblr}[
		caption={Summary of evasion attacks against machine learning-based malware detectors.},
		label={tab:evasion_attacks}
		]{
			width = \linewidth,
			colspec = {Q[127]Q[54]Q[133]Q[131]Q[115]Q[156]Q[202]},
			hline{1-10} = {-}{},
		}
		\textbf{Citation}        & \textbf{Year} & \textbf{Target Classifier}                & \textbf{Attack's Output}            & \textbf{Threat Model} & \textbf{Features}                      & \textbf{;anipulations}                                                                          \\
		Anderson et al.~\cite{DBLP:journals/corr/abs-1801-08917} & 2018 & {MalConv\\ LightGBM}             & PE file                    & Black-box    & Bytes                                   & {IAT\\ Overlay\\ Section injection\\ Section names rewritting\\ Packing}  \\
		Suciu et al.~\cite{DBLP:conf/sp/SuciuCJ19}    & 2019 & MalConv                          & PE file                    & White-box    & Bytes                                   & {Overlay\\ Slack space}                                                                   \\
		Demetrio et al.~\cite{DBLP:journals/tissec/DemetrioCBLAR21} & 2021 & {MalConv\\ AvastConv\\ LightGBM} & PE file                    & White-box \& Black-box   & Bytes                                   & {DOS header\\ PE headers}                                                                 \\
		Demetrio et al.~\cite{DBLP:journals/tissec/DemetrioCBLAR21} & 2021 & {MalConv\\ LightGBM}             & PE file                    & Black-box    & Bytes                                   & Section injection                                                                         \\
		Lucas et al.~\cite{DBLP:conf/asiaccs/LucasSBRS21}    & 2021 & {MalConv\\ AvastConv}            & PE file                    & Black-box    & Bytes                                   & Binary diversification                                                                    \\
		Hu et al.~\cite{DBLP:conf/dmbd/Hu022}       & 2022 & {API-based\\ detector}           & Feature vector             & Gray-box     & API functions and libraries                          & IAT                                                                      \\
		Yuste et al.~\cite{YUSTE2022102643}    & 2022 & MalConv                          & PE file                    & Black-box    & Bytes                                   & Slack space                                                                               \\
		Gibert et al.~\cite{gibert_evasion}   & 2023 & {MalConv\\ LightGBM}             & {Feature vector\\ PE file} & Black-box    & {Byte 1-Gram\\ API functions and libraries\\ Strings} & {IAT\\ Overlay}                                                     
	\end{talltblr}
\end{table}

Machine learning systems have been shown to be vulnerable to adversarial attacks, limiting the application of machine learning. This vulnerability is specially pronounced in a non-stationary and adversarial environment such as in the domain of malware detection, where actual adversaries, i.e. malware developers, exist and are incentivised to actively search for ways to circumvent detection. Although most of the research in adversarial learning has focused on the Computer Vision domain~\cite{DBLP:journals/corr/GoodfellowSS14,DBLP:conf/ccs/PapernotMGJCS17,DBLP:conf/sp/Carlini017,DBLP:conf/cvpr/EykholtEF0RXPKS18,DBLP:conf/icml/CohenRK19,DBLP:conf/nips/ShafahiNG0DSDTG19,DBLP:conf/iclr/WongRK20}, in recent years, adversarial example generation methods~\cite{DBLP:journals/csur/RosenbergSER21} have increasingly been explored to evade machine learning-based malware detectors, as summarized in Table~\ref{tab:evasion_attacks}. 

\begin{figure}[ht]
	\includegraphics[width=0.5\columnwidth]{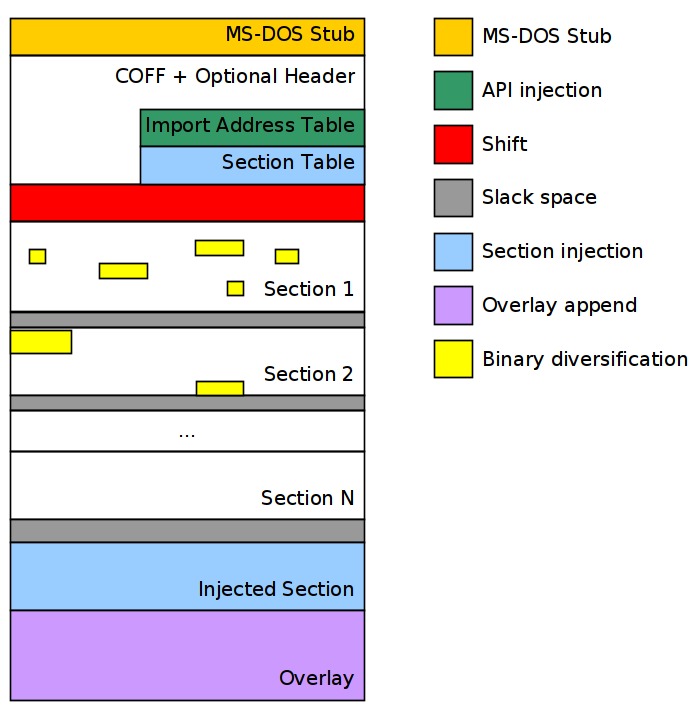}
	\centering
	\caption{A graphical depiction of the PE file format and some practical
		manipulations~\cite{DBLP:conf/sp/SuciuCJ19,DBLP:journals/tissec/DemetrioCBLAR21,demetrio2021functionality,YUSTE2022102643}.}
	\label{fig:pe_manipulations}
\end{figure}

While in some domains, like image classification, modifying one or more pixels in an image does not result in an incorrect image, modifying a single byte of a Portable Executable file might result in an unrunnable PE file. For this reason, researchers must be especially cautious in how they manipulate PE files to avoid breaking the functionality and format of the PE files. Nevertheless, there are a wide variety of ways in which PE files can be manipulated without rendering a corrupted executable, as summarized in Figure~\ref{fig:pe_manipulations}.

Hu et al.~\cite{DBLP:conf/dmbd/Hu022} presented MalGAN, a GAN-based approach to generate adversarial API-based feature vectors to attack static API-based malware detectors, i.e. a malware detector that only takes as input the API-based features extracted from the Import Address Table (IAT) of PE files. MalGAN consists of two feed-forward neural networks, a generator and a substitute detector. MalGAN works as follows: (1) the generator network is trained to minimize the generated adversarial malicious API-based feature vectors' maliciousness probabilities predicted by the substitute detector; (2) the substitute detector is trained to fit the target detection system. By training both networks together, the generator will learn what API libraries and functions have to add to the Portable Executable file to evade the target API-based malware detector.

\begin{figure}[ht]
	\includegraphics[width=\columnwidth]{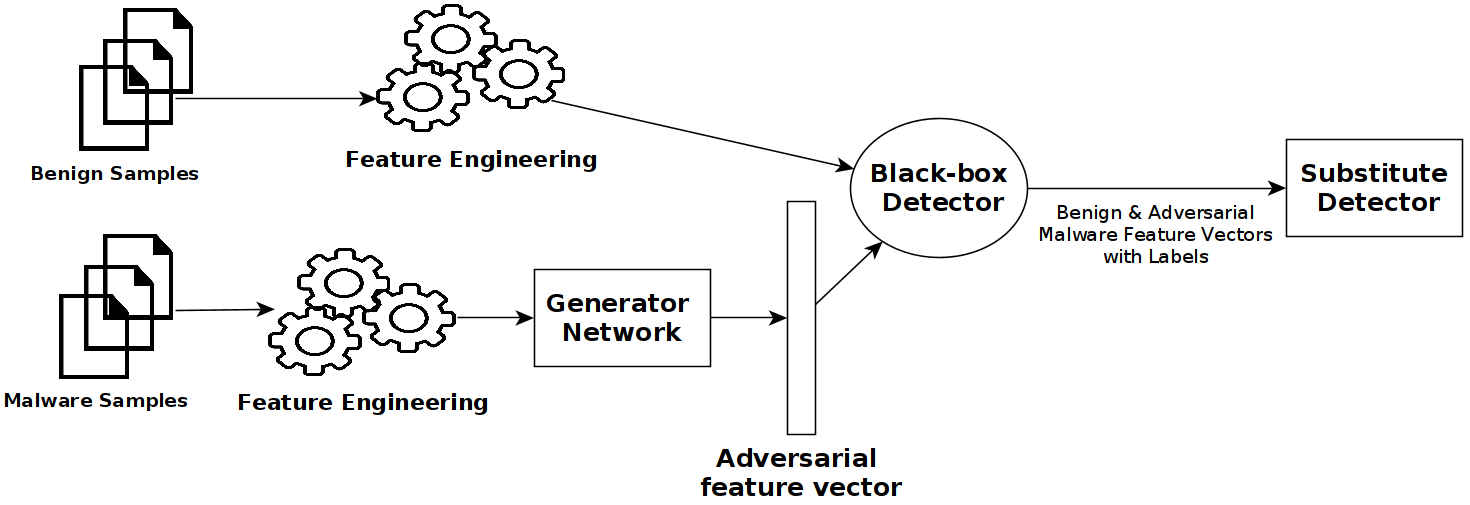}
	\centering
	\caption{Graphical depiction of MalGAN's architecture~\cite{DBLP:conf/dmbd/Hu022}.}
	\label{fig:malgan_architecture}
\end{figure}

\begin{figure}[ht]
	\includegraphics[width=0.7\columnwidth]{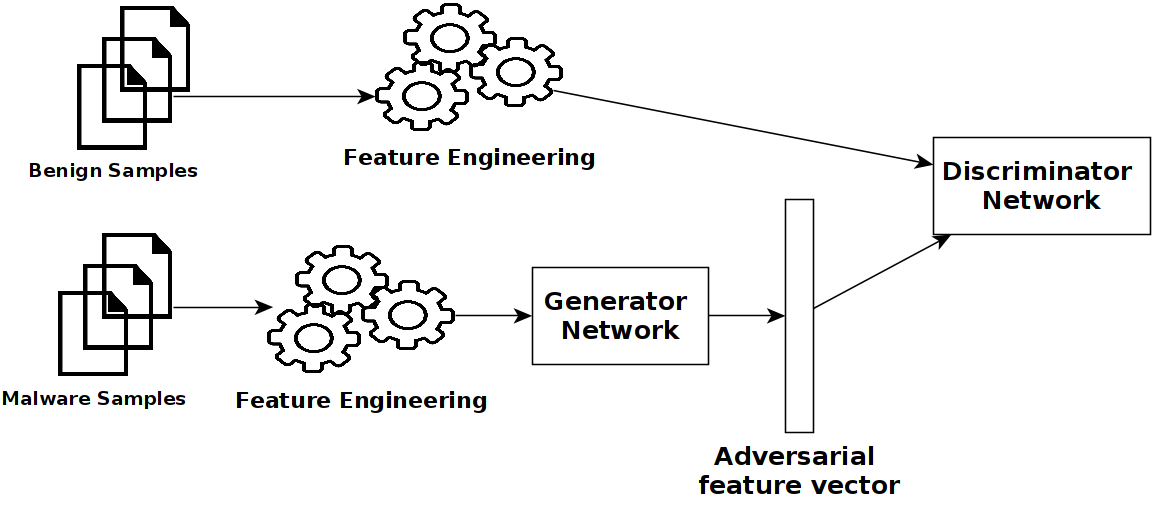}
	\centering
	\caption{Graphical depiction of the Wasserstein GAN architecture~\cite{gibert_evasion}}
	\label{fig:wasserstein_architecture}
\end{figure}

Gibert et al.~\cite{gibert_evasion} presented a GAN-based approach to generate adversarial feature vectors, including API-based features, byte 1-Gram features and String-based features, to attack SOTA malware detectors~\cite{DBLP:conf/aaai/RaffBSBCN18,2018arXiv180404637A} without having to query them. In their work, instead of training a substitute detector to emulate the target malware detector and then attacking the substitute detector, they proposed a general framework using a Conditional Wasstertein GAN to generate adversarial feature vectors that resemble benign feature vectors, thus fooling the malware detectors. The GAN-based architecture consists of a generator, which generates the adversarial feature vectors, and a critic network, which is trained to discriminate between benign and adversarial feature vectors. An overview of the proposed GAN-based approach is illustrated in Figure~\ref{fig:wasserstein_architecture}.

Suciu et al~\cite{DBLP:conf/sp/SuciuCJ19} proposed two attack strategies to manipulate and optimize the bytes of PE files using the Fast Gradient Sign method~\cite{DBLP:journals/corr/GoodfellowSS14} (FGSM): (1) appending and optimizing additional bytes at the end of the PE files, known as overlay append; (2) and manipulating the slack bytes, i.e. bytes inserted by the compiler between sections for alignment purposes. 

Demetrio et al.~\cite{DBLP:journals/tissec/DemetrioCBLAR21} presented a method for generating adversarial malware examples by manipulating the bytes in the DOS headers of PE files and by creating new space between the headers and the sections of PE files. The DOS header is primarily maintained for backward compatibility with older Windows operating systems and their bytes can be freely manipulated without breaking the functionality of the PE file.

Yuste et al.~\cite{YUSTE2022102643} proposed an attack strategy that consists of dynamically extending unused blocks between sections, referred to as code caves. Subsequently, their attack strategy allows for the injection of huge adversarial payloads between sections, which are optimized using a genetic algorithm.

Demetrio et al.~\cite{demetrio2021functionality} presented an evasion strategy that involves injecting benign content within newly-created sections. Afterwards, the injected adversarial payload is optimized using a genetic algorithm.  

Lucas et al.~\cite{DBLP:conf/asiaccs/LucasSBRS21} proposed an evasion attack that interweaves the following two binary-diversification techniques: in-place randomization and code displacement to produce adversarial malware examples. In-place randomization involves identifying functions and basic blocks and statically performing 4 types of operations: (1) replacing instructions with equivalent ones of the same length; (2) reassigning registers within functions or sets of basic blocks; (3) reordering instructions using a dependence graph; and (4) altering the order in which register values are pushed to and popped from the stack. Code displacement breaks potential portions of code by moving code to a new executable section with $jmp$ instructions.

Anderson et al.~\cite{DBLP:journals/corr/abs-1801-08917} proposed a general framework based on reinforcement learning (RL) for attacking static malware detectors. Within this framework, a RL agent, equipped with a set of functionality-preserving manipulations or actions, interacts with the target malware detector through a series of games until it learns which sequences of manipulations are more likely to result in evading the target malware detector. The actions of the agent are the following: (1) adding a function to the Import Address Table; (2) renaming existing section names; (3) creating new (unused sections; (4) appending bytes to the slack space; (5) appending bytes to the overlay; (6) removing signer information; (7) packing or unpacking the file; and (8) modifying (breaking) the header's checksum.

\section{Robust Machine Learning-based Malware Detectors}
Several strategies and techniques have been recently explored to defend against adversarial attacks and to build robust machine learning-based malware detectors, including adversarial training~\cite{287238}, eliminating attack vectors~\cite{DBLP:journals/corr/abs-2010-09569}, and smoothing-based defenses~\cite{gibert2023_randomizedsmoothing,huang2023rsdel,10.1145/3605764.3623914}.

\subsection{Adversarial Training}
Adversarial training is a machine learning technique used to improve the robustness of a model by exposing it to adversarial examples during training. To this end, Luca et al.~\cite{287238} investigated the effectiveness of using adversarial training to create robust end-to-end malware detectors against adversarial examples. In their work, they experimented with various strategies for augmenting the training data, including three state-of-the-art evasion methods: (1) in-Place replacement (IPR)~\cite{DBLP:conf/asiaccs/LucasSBRS21}, (2) displacement (Disp)~\cite{DBLP:conf/asiaccs/LucasSBRS21}, and (3) the attack proposed by Kreuk et al~\cite{DBLP:journals/corr/abs-1802-04528}. Moreover, they also experimented with random (unguided) transformations of the IPR and Disp attacks. The main findings of their work include the following:
\begin{itemize}
	\item Models can be made more robust to adversarial attacks by adversarially training them with lower-effort versions of the same attack.
	\item Adversarial training increases the robustness of models against the attacks seen during training (90\% $\rightarrow $ 5\%, 26\% $\rightarrow $ 6\%, and 84\% $\rightarrow $ 30\% attack success rate for Disp, IPR, Kreuk-based attacks, respectively), but are ineffective against state-of-the-art attacks.
\end{itemize}
In addition to being dependent on the adversarial attacks used during training, adversarial training increases the training time and it fails to provide a robustness certificate, limiting its applicability for the task of malware detection.

\begin{figure}[ht]
	\includegraphics[width=0.8\columnwidth]{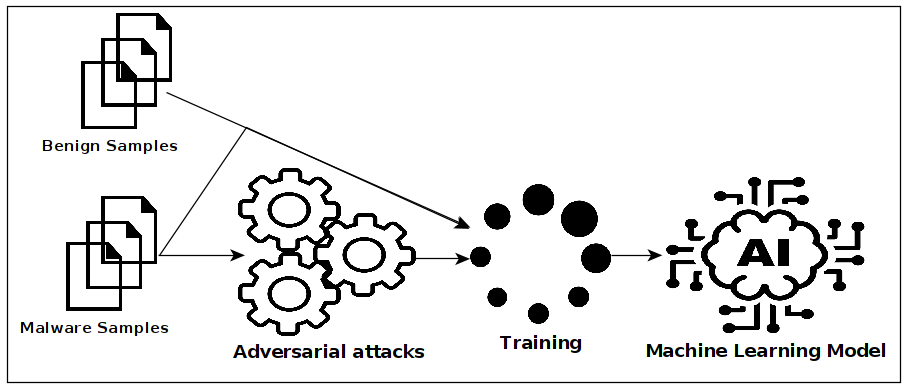}
	\centering
	\caption{Illustration of the adversarial training procedure.}
	\label{fig:adversarial_training_procedure}
\end{figure}

\subsection{Diversification and Elimination of Attack Vectors}
The use of machine learning for malware detection introduces new attack surfaces. Attackers started exploiting the format of Portable Executable files to change or add content in some parts to achieve evasion while keeping the functionality of the program intact~\cite{DBLP:conf/sp/SuciuCJ19,DBLP:journals/corr/abs-1802-04528,demetrio2021functionality,YUSTE2022102643,gibert_evasion}. As described in Section~\ref{sec:adversarial_attacks}, attackers can manipulate unused bytes in the DOS header~\cite{DBLP:journals/tissec/DemetrioCBLAR21}, append bytes to the overlay~\cite{DBLP:conf/sp/SuciuCJ19,DBLP:journals/corr/abs-1802-04528}, create new sections~\cite{demetrio2021functionality}, modify the bytes in the slack space at the end of each section~\cite{DBLP:conf/sp/SuciuCJ19}, and even enlarge the slack space between sections~\cite{YUSTE2022102643}.

A relatively straightforward, yet effective method to evade end-to-end malware detectors (See Section~\ref{sec:byte_based_detectors}) is to inject content extracted from benign executables into unused areas or into newly-created sections to overwhelm the malicious characteristics~\cite{demetrio2021functionality}. This approach has also been shown to be effective against feature-based detectors like the EMBER LighGBM model~\cite{2018arXiv180404637A}, as it modifies some of the features these detectors rely on, such as the strings, N-grams, the entropy of the file, and so forth. 

Given the facility with which attackers can manipulate Portable Executable files to inject new content, Quiring et al~\cite{DBLP:journals/corr/abs-2010-09569} proposed a malware detection system that combines multiple, diverse defenses, including a "file format" detector, named "Semantic Gap" in their work. This "file format" detector scans the overlay, the slack space, and the sections to check if an attacker has exploited the file format to inject an adversarial payload. The slack space scanner scans the slack space to detect if the space between sections is filled with non-zero bytes. The overlay scanner checks if the attacker has appended bytes at the overlay by computing the ratio of the overlay of the overall file size. Finally, the section scanner checks if there are two or more duplicate sections. This scanner has been implemented based on the assumption that benign files have rather unique sections and that only the malicious executables will contain duplicate sections. The goal of these scanners is to force the attacker to perform more complicated changes than adding content in unused, well-known areas. 

\begin{figure}[ht]
	\includegraphics[width=1.0\columnwidth]{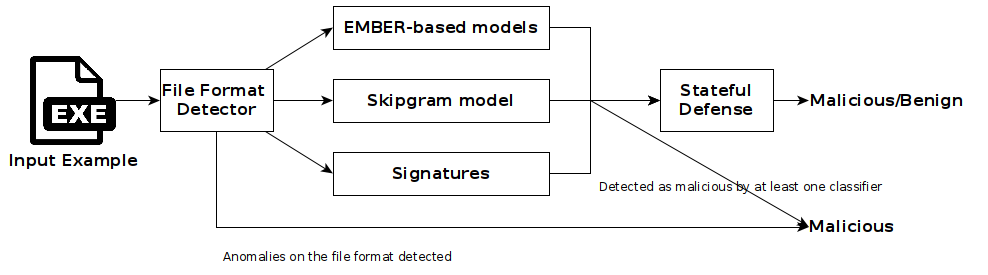}
	\centering
	\caption{Overview of the diversification-based defense presented by Quiring et al.~\cite{DBLP:journals/corr/abs-2010-09569}.}
	\label{fig:diversification_defense}
\end{figure}

Following the "file format" detector, the malware detection system proposed by Quiring et al.~\cite{DBLP:journals/corr/abs-2010-09569} combines various types of classifiers including four EMBER-based GBDT models, a Skipgram model, and a signature-based model, and a stateful defense. In their system, a PE file is classified as malware if any of the system's components consider it malicious. The main idea is to combine multiple detectors to increase the likelihood of correctly identifying malware, forcing the attacker to exploit the weaknesses in all components in order to achieve evasion. The overall system is presented in Figure~\ref{fig:diversification_defense}. Table~\ref{tab:diversification_defense} provides a description of all the components in their approach.

\begin{table}[ht]
	\centering
	\begin{talltblr}[
		caption={Description of the components of the diversification-based defense proposed by Quiring et al.\cite{DBLP:journals/corr/abs-2010-09569}.},
		label={tab:diversification_defense}
		]{
			width = \linewidth,
			colspec = {Q[148]Q[213]Q[579]},
			cell{2}{1} = {r=3}{},
			cell{5}{1} = {r=6}{},
			cell{5}{2} = {r=4}{},
			hline{1-2,5,11-12} = {-}{},
			hline{9-10} = {2-3}{},
		}
		Defense              & Type                    & Description                                                                           \\
		File format detector & Slack space             & Scans the slack space to detect if the space between sections is filled with non-zero bytes.                                                          \\
		& Overlay                 & Checks if the overlay is too large compared to the file size.                               \\
		& Section duplicates      & Checks if duplicate sections are used.                                                       \\
		Malware detectors       & EMBER-based GBDT models & Default: GBDT model without PE header features, trained on 2017 corpus.                           \\
		&                         & 1st variation: GBDT model without PE header, string stream removed, input truncated, 2017 corpus. \\
		&                         & 2nd variation: GBDT model with reduced feature set, input truncated, 2017 corpus.                      \\
		&                         & 3rd variation: GBDT model with reduced feature set, input truncated, 2018 corpus.                      \\
		& Skipgram model          & Monotonic GBDT model based on skipgrams (3-skip 3-grams).                             \\
		& Signature-based model   & Detection based on Yara rules for popular malware families.                            \\
		Stateful defense     &                         & Monitors incoming queries for patterns indicative of evasion.                           
	\end{talltblr}
\end{table}

\subsection{Smoothing-based defenses}
\begin{figure}[ht]
	\includegraphics[width=1.0\columnwidth]{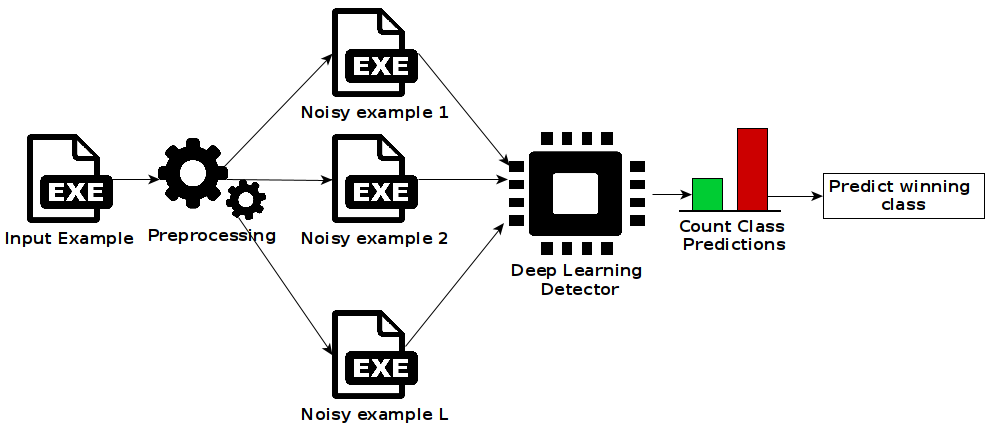}
	\centering
	\caption{Graphical depiction of the randomized smoothing scheme for malware detection~\cite{gibert2023_randomizedsmoothing,huang2023rsdel}.}
	\label{fig:randomized_smoothing_scheme}
\end{figure}

Smoothing-based defenses are a category of techniques used in the field of adversarial machine learning. The idea behind smoothing-based defenses is to apply a form of preprocessing or transformation to the input data in order to make the model more robust against adversarial perturbations. Smoothing-based defenses work as follows: (1) at training time, a base classifier is trained to make classifications based on a noisy version of a given input example; (2) at inference time, $L$ noisy version of an input example are generated, independently classified, and the final classification is determined as the class that the classifier most frequently predicts across the set of noisy versions of the input example. An overview of the smoothing-based defense strategy is presented in Figure~\ref{fig:randomized_smoothing_scheme}. The key idea behind smoothing-based defenses involves adding a careful amount of noise or adversarial perturbation to the input data and then observing the statistical distribution of the model's predictions over multiple noisy samples. By doing so, the model's decision boundaries become more diffuse, making it less sensitive to small perturbations in the input~\cite{DBLP:conf/icml/CohenRK19}.

In the context of malware detection, two randomized smoothing-based defenses, illustrated in Table~\ref{tab:randomized_smoothing_schemes_graphical_depiction}, have been proposed so far:
\begin{itemize}
	\item Randomized ablation-based smoothing~\cite{gibert2023_randomizedsmoothing}. This strategy replaces a random subset of bytes from the input data with a special NULL value.
	\item Randomized deletion-based smoothing~\cite{huang2023rsdel}. This strategy applies random byte deletions to the input data. 
\end{itemize}
Both strategies show greater robustness compared to a baseline end-to-end classifier (MalConv model) against various state-of-the-art attacks in the literature at the expense of increased computational time.

\begin{table}
	\subcaptionbox{Original file}{
		\begin{tabular}{cc}
			\hline
			File offset & Byte                        \\ \hline
			00000000    & 77  \\
			00000001    & \cellcolor[HTML]{C0C0C0}90  \\
			00000002    & 144                         \\
			...         & ...                         \\
			00000400    & 85                          \\
			00000401    & \cellcolor[HTML]{C0C0C0}139 \\
			00000402    & \cellcolor[HTML]{C0C0C0}236 \\
			00000403    & 131                         \\
			00000404    & 236                         \\
			00000405    & \cellcolor[HTML]{C0C0C0}92  \\ \hline
		\end{tabular}%
	}
	\hfill
	\subcaptionbox{File under randomized ablation-based scheme.}{
		\begin{tabular}{cc}
			\hline
			File offset & Byte \\ \hline
			00000000    & 77   \\
			00000001    & NULL   \\
			00000002    & 144  \\
			...         & ...  \\
			00000400    & 85   \\
			00000401    & NULL  \\
			00000402    & NULL  \\
			00000403    & 131  \\
			00000404    & 236  \\
			00000405    & NULL   \\ \hline
		\end{tabular}%
	}
	\hfill
	\subcaptionbox{File under randomized deletion-based scheme.}{
		\begin{tabular}{cc}
			\hline
			File offset & Byte \\ \hline
			00000000    & 77   \\
			00000002    & 144  \\
			...         & ...  \\
			00000400    & 85   \\
			00000403    & 131  \\
			00000404    & 236  \\\hline
		\end{tabular}%
	}
	\hfill
	\caption{Illustration of the ablation-based and deletion-based smoothing mechanisms applied to an executable file. Left: An executable file where the byte sequence representation is shown in the 2nd column. Shading represents bytes that are ablated or deleted in the corresponding perturbed file. Middle: A perturbed file produced by the ablation mechanism. Right: A perturbed file produced by the deletion mechanism.}
	\label{tab:randomized_smoothing_schemes_graphical_depiction}
\end{table}

\subsubsection{(De)Randomized Smoothing for Malware Detection}
Randomized smoothing, however, while it confers some robustness against some types of attacks~\cite{DBLP:conf/sp/SuciuCJ19,DBLP:conf/asiaccs/LucasSBRS21,DBLP:journals/tissec/DemetrioCBLAR21}, it has been shown to be ineffective in defending against adversarial malware examples generated
with state-of-the-art evasion attacks~\cite{demetrio2021functionality, YUSTE2022102643}. Randomized smoothing-based defenses create noisy examples by randomly replacing~\cite{gibert2023_randomizedsmoothing} or deleting bytes~\cite{huang2023rsdel}, which greatly differ from the types of manipulations that attackers employ to evade machine learning-based detectors. 

In the context of malware detection, the bytes within executable files serve functional purposes and cannot be arbitrarily manipulated without potentially rendering the executable nonfunctional. Unlike the manipulation flexibility found in the image
domain, where attackers can alter any pixel in an image at will,
indiscriminate modifications to the bytes in a file can disrupt
the intended functionality of the executable. To circumvent this challenge, attackers commonly inject adversarial payloads into specific locations of the executable files.

Due to this distinct nature of adversarial attacks in the realm
of malware, where adversarial payloads are injected into specific
locations of files, the application of randomized smoothing,
employing random noise, may not be the most suitable defense
mechanism against adversarial malware examples.

\begin{figure}[ht]
	\includegraphics[width=0.8\columnwidth]{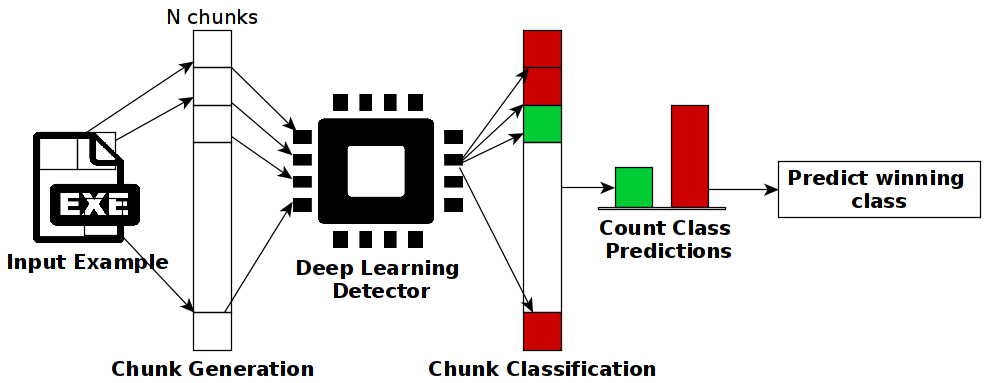}
	\centering
	\caption{Graphical depiction of the (de)randomized smoothing scheme for malware detection~\cite{10.1145/3605764.3623914}.}
	\label{fig:derandomized_smoothing}
\end{figure}

To overcome the limitations of randomized smoothing, Gibert et al.~\cite{10.1145/3605764.3623914} proposed a smoothing-based defense inspired by (de)randomized smoothing~\cite{DBLP:conf/nips/0001F20a}, a class of certified robust image classifiers designed to defend against patch attacks. In their work, Gibert et al~\cite{10.1145/3605764.3623914} adapted the concept of (de)randomized smoothing to detect malware. Their proposed defense, instead of employing adding random noise, removes whole subsequences of contiguous bytes. Figure~\ref{fig:derandomized_smoothing} illustrates the (de)randomized smoothing approach for malware detection. 

Their (de)randomized smoothing scheme works as follows: 
\begin{enumerate}
	\item At training time, a base malware classifier is trained to make classification based on ablated versions of the input examples. These ablated versions consist of only a subset of contiguous bytes or a chunk of bytes from a randomly selected location from the original input examples. The rest of the file is removed, i.e. it is not used for classification.
	\item At inference time, a given input example is divided into non-overlapping chunks, each chunk is classified independently, and then, the final classification output is computed using a majority vote over all the chunk predictions.
\end{enumerate}
By leveraging the fact that the adversarial payload injected by attackers to evade detection can only influence a certain number of chunks, they derived deterministic robustness certificates against patch and append attacks~\cite{DBLP:journals/tissec/DemetrioCBLAR21,DBLP:conf/sp/SuciuCJ19} if the number of predictions of the correct class exceeds the second most commonly predicted class by a large enough margin.

\section{Conclusion}
In this chapter, we have explored how machine learning is being used to build malware detection systems for the Windows operating system, encompassing both feature-based, deep learning-based detectors, and visualization techniques to aid analysts. However, while machine learning approaches have been shown to be very valuable for complementing traditional signature-based and heuristic-based detection methods, this chapter underscores the importance of tackling the inherent challenges of these detectors. Addressing concept drift and the robustness against adversarial attacks is crucial for building robust malware detectors due to the dynamic and evolving nature of both the threat landscape, with the appearance of new malware families and variants, and potential evasion techniques employed by malicious actors, which render ML-based malware detectors ineffective. To this end, ongoing research on improving and developing adversarial defenses is crucial to maintain the effectiveness of these detectors in the face of evolving threats, adversarial attacks, and changes in the characteristics of malware.

\section*{Acknowledgments}
This work has received funding from Enterprise Ireland and the European Union’s Horizon 2020 Research and Innovation Programme under the Marie Skłodowska-Curie grant agreement No 847402. 

\bibliographystyle{unsrt}  
\bibliography{references.bib}

\end{document}